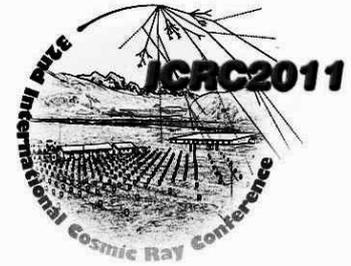

# Implementation of the readout system in the UFFO Slewing Mirror Telescope

J.E. KIM[1], H.LIM[1], A. JUNG[1], K.-B AHN[2], H.S. CHOI[3], Y.J. CHOI[4], B. GROSSAN[5], I. HERMANN[4], S. JEONG[1], S.-W. KIM[2], Y.W. KIM[1], J. LEE[1], E. V. LINDER[1,5], K.W. MIN[4], G.W. NA[1], J.W. NAM[1], K.H. NAM[1], M.I. PANAYUK[6], I. H. PARK[1], G. F. SMOOT[1,5], Y.D. SUH[4], S. SVELITOV[6], N. VEDENKEN[6], I. YASHIN[6], M.H. ZHAO[1], FOR THE UFFO COLLABORATION
[1]*Ewha Womans University, Seoul, Korea*
[2]*Yonsei University, Seoul, Korea*
[3]*Korea Institute of Industrial Technology, Ansan, Korea*
[4]*Korea Advanced Institute of Science and Technology, Daejeon, Korea*
[5]*University of California, Berkeley, USA*
[6]*Moscow State University, Moscow, Russia*
jekim@hess.ewha.ac.kr

**Abstract:** The Ultra-Fast Flash Observatory (UFFO) is a new space-based experiment to observe Gamma-Ray Bursts (GRBs). GRBs are the most luminous electromagnetic events in the universe and occur randomly in any direction. Therefore the UFFO consists of two telescopes; UFFO Burst Alert & Trigger Telescope (UBAT) to detect GRBs using a wide field-of-view (FOV), and a Slewing Mirror Telescope (SMT) to observe UV/optical events rapidly within the narrow, targeted FOV. The SMT is a Ritchey-Chretien telescope that uses a motorized mirror system and an Intensified Charge-Coupled Device (ICCD). When the GRB is triggered by the UBAT, the SMT receives the position information and rapidly tilts the mirror to the target. The ICCD start to take the data within a second after GRB is triggered. Here we give the details about the SMT readout electronics that deliver the data.

## 1. Introduction

The Ultra-Fast Flash Observatory (UFFO) is a space-based experiment to observe the prompt optical/UV photons from Gamma-Ray Bursts (GRBs). The UFFO pathfinder, the pilot experiment of UFFO, will be launched into space in November 2011 by the *Lomonosov* spacecraft. Its purpose and concept have been described in [1]. It consists of two instruments: UFFO Burst Alert & Trigger Telescope (UBAT) for measurement of GRB location and Slewing Mirror Telescope (SMT) for detection of UV/optical events.

The main concept of SMT is very fast pointing of the narrow field-of-view (FOV) using a fast steerable mirror plate. When the UBAT triggers GRBs, it sends SMT the trigger signal and position information. The SMT calculates how many degrees the mirror is to be tilted and starts to take the data using the Intensified Charge-Coupled Device (ICCD). This process is done within seconds after a GRB is triggered and consequently the UFFO can detect the early emission from GRBs. In this paper, we describe our approach and the current status of the SMT readout system.

## 2. Overview of SMT

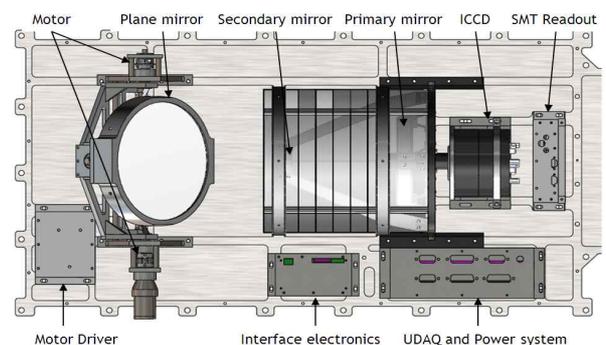
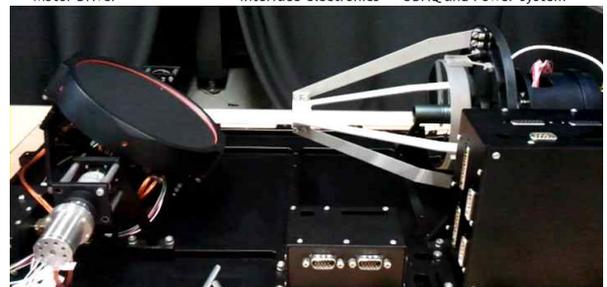

Figure 1. SMT Instrument

The SMT is a Ritchey-Chretien telescope (Figure 1) with a 100 mm diameter aperture using a motorized mirror plate and with an f-number of 11.4 [3]. The detector is an ICCD with Micro-Channel Plates (MCPs); it can observe faint events such as a 19.11 magnitude B-star in white light per 100ms and operate in photon counting mode. The detector FOV is 17 x 17 arcmin$^2$, and each pixel has a size of 4 x 4 arcsec$^2$. The wavelength range is 200-650 nm. The mirror rotation angle is ±17.5°, resulting in an accessible FOV of 60° x 60°, without the aberration inherent in wide-field optical systems [1]. The SMT has the readout rate of 4 ms and can take 250 frames per second. It takes only 1s to receive the trigger signal to slew the motorized mirror forward to target and to take UV/Optical data from target. A brief specification of SMT can be found in Table 1.

| Telescope | Ritchey-Chrétien + Motorized mirror plate |
|---|---|
| Aperture | 100 mm diameter |
| F-number | 11.4 |
| Detector | Intensified CCD with MCPs |
| Detector Operation | Photon Counting |
| Field of view | 17 x 17 arcmin |
| Number of Pixels | 256 x 256 pixels |
| Pixel FOV | 4 arcsec |
| Wavelength Range | 200 nm ~ 650 nm |
| Readout Rate | 4 ms |
| Data rate | 150 Mbyte/day |
| Processing time | ~1 s |

Table 1. SMT Specification

## 3. Implementation of SMT readout system

### 3.1 Purpose and goal

The main goal of the SMT readout system is the data acquisition and storage. Also, the readout system creates the clocks to control the CCD, calculates an angle to slew the mirror, drives the motor, communicates with an interface instrument, controls the voltage of ICCD to optimize the gain, and monitors the housekeeping data such as temperature and current. We designed the circuits and the trigger/control/data flows for our approach and built the readout system including a detector.

### 3.2 Detector

The detector of SMT is an ICCD that consists of an intensifier and a CCD readout system. These are coupled with the fiber optics taper. The intensifier is comprised of an input window, an UV-enhanced S20 photocathode, two stacked MCPs, and a phosphor screen as shown Figure 2. The intensifier is optimized over 200-650 nm wavelength, determined by the photocathode and the detector input window. Photons entering ICCD are converted to electrons at the photocathode. These electrons are multiplied at the MCPs, and clusters of these electrons strike the phosphor screen and change the amplified electrons back into visible light. This light is focused onto the CCD camera through the fiber optic taper.

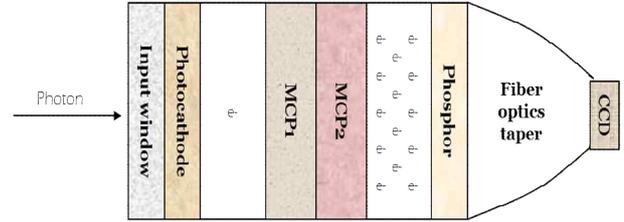

Figure 2. ICCD structure

The UV-enhanced S20 photocathode provides the better sensitivity for the UV/optical wavelength (< 350 nm) and has a dark current of about 500 e-/cm2/s at room temperature. An electron from the photocathode enters the MCP, and produces the secondary electrons when it strikes the inner wall of the MCP. These electrons are further accelerated by a voltage applied across the MCP to produce additional secondary electrons. The electron multiplication factor or gain per MCP is an average of 1,000 times. The double MCPs are used in the SMT; the ICCD has the gain up to $10^6$. The noise contribution is more critical for the long exposure time so that the double MCPs are the better solution for our application.

For the phosphor screen, the P46 material with the decay time of 300 ns and with light emission from 490 nm to 620 nm is used. The phosphor with the fast decay time is required due to the fast readout system. Light from the phosphor screen is passed through a fiber taper and the CCD used to detect the photons. The sensor is a Kodak KAI-0340 which is a commercial interline CCD. It provides a fast pixel-readout up to 40 MHz. The CCD has 640 x 480 pixels; however, 256 x 256 pixels will be used for our observations. The coupling between the MCPs and CCD is in the ratio of 3:1, and thus the total focal plane size of ICCD is 15.155 x 11.366 mm$^2$. The actual focal plane size of SMT will be 5.606 x 5.606 mm$^2$ with 256 x 256 pixels, and the pixel size is 22.2 x 22.2 μm$^2$.

### 3.2 Readout/Control architecture

The SMT readout system is controlled by the Field-Programmable Gate Array (FPGA) chip, for which we chose A3P1000-PQ208 manufactured by ACTEL. It consists of several readout and control units as shown Figure 3.

To control the SMT, we employed the UFFO Data Acquisition (UDAQ) system, an interface instrument, which is described in [4]. The SMT receives trigger and control signals from the UDAQ, and also sends the data. The communication between the SMT and the UDAQ is based on a Serial Peripheral Interface bus (SPI). The Internal Interface Unit (IIU) in the SMT converts the signal type differently defined by each sub-system into the proper type.

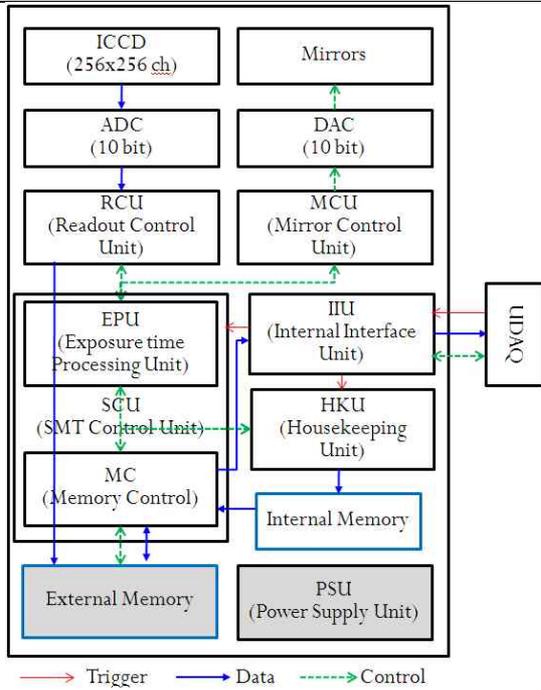

Figure 3. SMT Readout/Control Architecture

After the UFFO power turns on, the UDAQ sends SMT the acknowledgement of initial state. The SCU turns on the power of the motor system and makes the motor be initialized and stabilized. Then it turns on the power of ICCD and makes for the ICCD to control and monitor the gain up to $10^6$. The Exposure time Processing Unit (EPU) sets the exposure time to the initial value and the Static Random Access Memory (SRAM) is ready to take the data though the Memory Control (MC).

When a GRB is triggered by UBAT, the trigger signal and coordinate information are passed to Mirror Control Unit (MCU) of SMT through UDAQ. The MCU sends the motorized mirror the information, and then the exposure time is set by EPU. When the motorized mirror is slewed onto the target, the SMT is ready to take the data.

The CCD needs 14 kinds of clock in order to be operated. We developed the control clocks in the Readout Control Unit (RCU). A frame consists of 256 vertical clocks and one vertical clock has 256 horizontal clocks. Figure 4 shows the detail information.

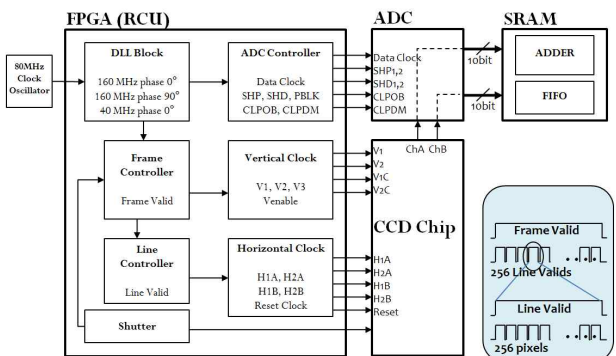

Figure 4. CCD control clocks of RCU

The UFFO has a limit in the data size, 300 Mbytes a day. Half is allotted for the SMT and half is for the UBAT so that SMT can take 150 Mbytes a day.

The ICCD is read out in 10 bit digitized values using ADCs. The pixel readout rate is 40 MHz and the number of readout pixels is 256 x 256. It takes about 2 ms to take the data of a frame and the exposure time is also 2 ms. Thus the ICCD readout time is 4 ms. The number of frames will be digital-summed in FPGA for a given exposure time that is predefined in each orbit. The readout dynamic range when summing over exposure time is 8 bits. Thus, the size of a frame by digital summation is 540 kbit per frame. An event is made of about 800 frames, and consequently the size of raw data is about 54 Mbytes/event (= 540 kbits/frame x 800 frames / 8bits).

The GRB event is detected in a pixel of ICCD, so that an event size can be reduced by factor of about 10. We are developing the method of zero suppression with FPGA. We can save the storage space and take the more triggered events through this method. In summary, the final event size of SMT is 5 Mbytes per event, and we expect at least 30 triggers transferred to the ground every day.

### 3.3 Status of readout electronics

The engineering model of SMT electronics has been produced. It consists of 4 parts; a CCD board to mount the CCD chip, a Clock generator board to control the ICCD, a SMT DAQ board for data acquisition and system control, and a motor control system. Figure 5 shows the readout electronics boards. The power is supplied by UFFO power system, and we expect the power consumption to be 10W.

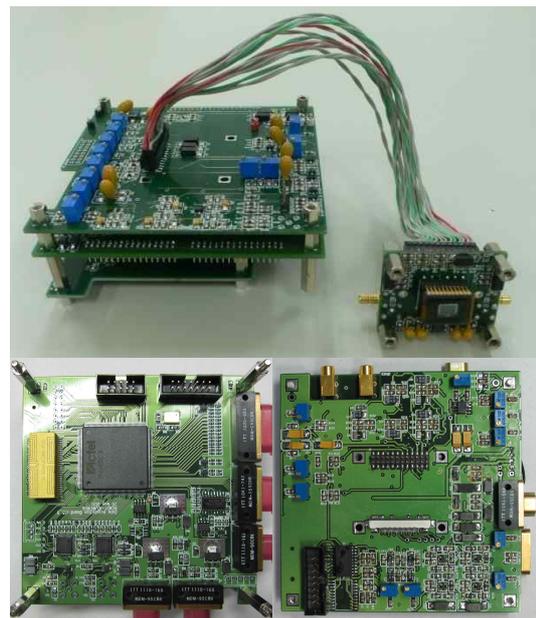

Figure 5. SMT Readout electronics assembly (Top), DAQ board (Left), Clock generator board (Right)

The CCD is controlled by the clock generator board with the specified performance at 40 MHz. It consists 4 hori-zontal clock drivers, 4 vertical clock drivers, a reset clock driver, 2 area selection circuits, and an electrostatic discharge (ESD) protection. The clocks and signals from this board are transmitted to the CCD board by the cables. The output signal of the CCD is buffered at the CCD board and the DAQ board, so that the signal passes between two boards without the loss of current. DAQ board has the gain control circuit for the high voltage of ICCD, the interface circuits for the motorized mirror, two Analog Front-Ends (AFEs) processing the analog output signal of CCD, a SRAM, and housekeeping (Figure 6).

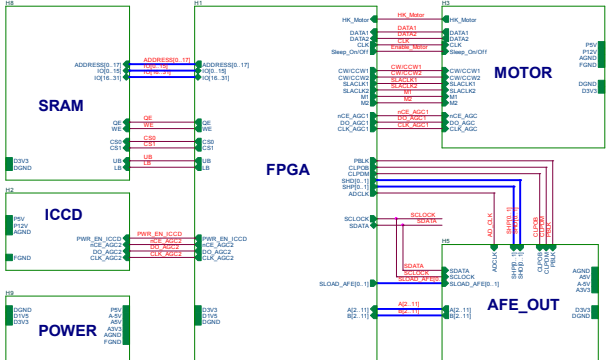

Figure 6. Block Diagram of SMT DAQ

We tested the readout part without the intensifier. For the lab test, the SMT electronics is interfaced by a USB-8451 based on a SPI communication and controlled by the standalone software which we designed in LABVIEW (Figure 7). We also used a commercial lens to focus on a near field image. Figure 8 shows the image that was taken by the SMT readout system. We had checked the implementation of electronics and logic chain in the lab test.

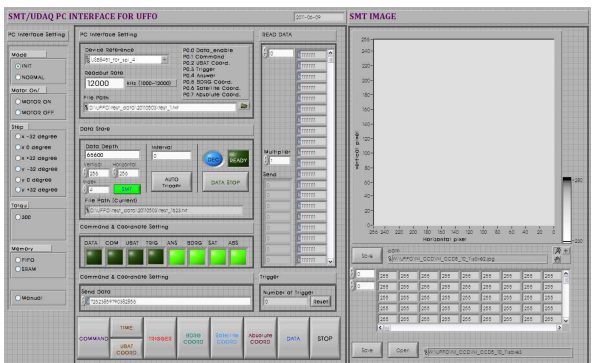

Figure 7. Standalone software for lab-test

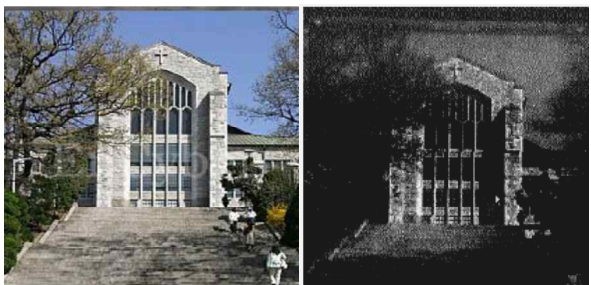

Figure 8. Target image (Left), CCD image (Right)

## 4. Summary

The UFFO is a space mission to detect early photons from GRBs at the time scale of sub-second for the first time. The SMT, one of the telescopes of UFFO, is prepared with the hardware including the optic, the detector and readout electronics. We have the space environment test ahead. The software parts are also being developed and updated for high performance such as trigger logic, data reconstruction, and so on. We expect at least 30 triggers transferred to the ground every day.

[1] I.H. Park et al., arXiv:0912.0773
[2] J.S. Bloom et al., Astrophysical Journal, 2009, 691: 723–737
[3] S. Jeong et al., Proceedings of this conference, #1269
[4] G.W. Na et al., Proceedings of this conference, #1258